\documentclass[12pt,pra,aps]{revtex4}
\usepackage{graphicx,caption}
\usepackage{array}
\usepackage{multirow}
\usepackage{epstopdf}
\usepackage{upgreek}
\usepackage{amsmath,amsfonts,amsthm,bm} 
\usepackage{setspace}
\usepackage{braket}
\usepackage{color}
\usepackage{booktabs}
\usepackage{hyperref}
\usepackage{enumerate}
\usepackage{soul}
\setcitestyle{super}
\usepackage{comment}
\usepackage{caption} 
\begin{document}

\title{Signatures of s-wave scattering in bound electronic states}
\author{Robin E. Moorby,$^{\ast a,b}$ Valentina Parravicini,$^a$ 
Maristella Alessio,$^a$ and Thomas-C. Jagau$^{\ast a}$\\
{\small $^a$Department of Chemistry, KU Leuven, Celestijnenlaan 200F, B-3001 Leuven, Belgium}\\
{\small $^b$Department of Chemistry, University of Durham, South Road, Durham, DH1 3LE, United Kingdom}\\
$^{\ast}$ Corresponding author: robin.moorby@kuleuven.be; thomas.jagau@kuleuven.be}

\begin{abstract}

\noindent
{\bf Abstract}

\noindent
We compute EOM-EA-CCSD and EOM-EA-CCSDT potential energy curves and one-electron 
properties of several anions at bond lengths close to where these states become 
unbound. In the potential energy curves of the totally symmetric anions of HCl 
and pyrrole, which are associated with s-wave scattering states at the equilibrium 
bond lengths of the parent neutral molecules, we observe on inclusion of diffuse 
basis functions a pronounced bending effect near the crossing points with the 
potential energy curves of the neutral molecules. Additionally, we observe that 
the Dyson orbital and second moment of the electron density become extremely 
large in this region. In particular, the second moment of the HCl anion becomes 
5 orders of magnitude times larger over a range of 5 pm. This behaviour is very 
different to the well-characterised non-totally symmetric anions of N$_2$ and 
H$_2$ that correspond to electronic resonances at the equilibrium bond lengths 
of their parent neutral molecules. Our work thus shows that bound state 
electronic-structure methods can distinguish between anions that turn into 
electronic resonances and those associated with s-wave scattering states.
\end{abstract}

\maketitle

\section{Introduction}

The attachment of electrons to neutral molecules can lead to bound or unbound 
molecular anions. If the electron affinity of a molecule is positive, the 
attachment of the excess electron is permanent, and the anions can be treated 
theoretically with conventional Hermitian quantum-chemical methods. One can 
distinguish between valence anions, in which the additional electron resides 
in a rather compact orbital close to the molecule, and non-valence anions, 
where the additional electron resides in a diffuse orbital and is bound 
by either the molecular dipole moment, quadrupole moment, or dispersion 
effects.\cite{Simons2008,Herbert2015,Jordan2017,Rogers2018,Simons2023,
Paran2024} For the computational treatment of non-valence anions in particular, 
large basis sets are mandatory to capture the enormous extent of the wave 
function. Also, correlated wave function theory is often necessary to compute 
accurate electron affinities and, in some cases, an anion is not bound at 
all at the Hartree-Fock (HF) level of theory. 

In cases where the anion is unbound at the equilibrium geometry of the 
neutral molecule, a non-zero angular momentum of the incoming electron 
can introduce a centrifugal barrier behind which the electron is temporarily 
trapped allowing for the formation of metastable electronic states, known 
as shape resonances.\cite{McVoy1968,Taylor1972,Simons2008} The energy of 
electronic resonances places them within a continuum of elastically scattered 
states, meaning that they can decay via electron loss due to their coupling 
with the continuum.\cite{Moiseyev2011,Jagau2017,Jagau2022} Such shape 
resonances are widespread, two well-characterized examples are the anions 
of H$_2$ and N$_2$. Besides shape resonances, there are Feshbach resonances 
which decay by a two electron process. A simple case of this type of state 
is the $(\upsigma_\mathrm{g})^1 (\upsigma_\mathrm{u})^2$ excited state of H$_2^-$ at stretched 
bond lengths.\cite{Bardsley1978,Stibbe1998,White2017} 

In the case of s-wave scattering, however, there is no centrifugal barrier. 
Temporary electron-attached states near the continuum threshold have 
been termed virtual states in s-wave scattering and are distinct from 
resonances.\cite{McVoy1968,Taylor1972,Domcke1991,Hotop2003} Virtual 
states have been associated with anions in molecules such as CO$_2$,\cite{
Morrison1982,Sommerfeld2003,Dvorak2022a,Dvorak2022b} (CN)$_2$,\cite{Nag2020} 
Fe(CO)$_5$,\cite{Allan2018}, HCl\cite{Nesbet1977,Cizek1999,Fedor2010} and 
pyrrole.\cite{Ragesh2022}

The asymptotic behaviour of unbound electronic states means that they cannot 
be treated with bound-state quantum chemical methods which fail to impose 
the correct boundary restrictions. For electronic resonances, non-Hermitian 
quantum chemistry offers an elegant solution to this problem: the energies 
are calculated as complex-valued eigenvalues of a non-Hermitian 
Hamiltonian.\cite{Moiseyev2011,Jagau2017,Jagau2022} Methods in this regard 
include complex scaling,\cite{Aguilar1971,Balslev1971,Simon1972} where the 
coordinates of the Hamiltonian are rotated into the complex plane; complex 
basis functions, where the exponents of basis functions are rotated into the 
complex plane;\cite{McCurdy1978} and complex absorbing potentials,\cite{
Jolicard1985,Riss1993} where a complex potential is applied to absorb the 
tails of the wave function. 

These techniques can be integrated into bound-state quantum chemistry 
enabling a treatment of electronic resonances in analogy to that of 
bound states. However, their applicability to s-wave scattering and 
virtual states is questionable, which means that one must resort to 
more traditional methods based on scattering theory.\cite{Taylor1972} 
The combination of the latter approaches with bound-state methods is 
still very challenging, even for diatomics such as HCl,\cite{Domcke1985,
Allan2000,Cizek1999,Fedor2010} which often necessitates the use of 
relatively low-level electronic-structure methods. 

It is thus desirable to identify and characterize anions that correspond 
to virtual states and distinguish them from other anions that correspond 
to electronic resonances. To this end, we investigate in this work the 
anions of HCl and pyrrole both of which are relevant to s-wave scattering 
and contrast them with H$_2^-$ and N$_2^-$ as prototypical examples of 
resonances. While these anions all are unbound at the equilibrium structures 
of the respective neutral molecules, they become bound upon bond stretching. 
We show in this work that close to the point at which an anion becomes 
unbound, bound states turning into resonances and s-wave scattered states, 
respectively, exhibit different behaviour in terms of potential energy 
curves and one-electron properties. 

\section{Computational Details}

We conduct coupled-cluster singles and doubles (CCSD) calculations\cite{
Purvis1982,Shavitt2009} on the ground states of the neutral HCl, N$_2$, 
H$_2$, and pyrrole molecules and equation-of-motion electron-attachment 
CCSD (EOM-EA-CCSD) calculations\cite{Stanton1993,Nooijen1995} to access 
the corresponding anions. We construct customized basis sets from the 
aug-cc-pVTZ basis\cite{Dunning1989,Kendall1992,Woon1993} with additional 
diffuse functions obtained by recursively dividing the exponent of the 
most diffuse s and p shell by a factor of 2. A basis set labelled 
aug-cc-pVTZ+$n$s$n$p indicates that we have added $n$ additional s and p 
shells to all atoms according to this even-tempered manner. In the largest 
basis sets used, aug-cc-pVTZ+15s15p, the smallest exponent, corresponding 
to an s function on hydrogen, is of the order of 10$^{-7}$. The molecular 
geometry of pyrrole is taken from Ref. \citenum{Mukherjee2022}.

For all systems but pyrrole, we also carry out coupled-cluster singles, 
doubles and triples (CCSDT)\cite{Noga1987,Noga1988} and equation-of-motion 
electron-attachment CCSDT (EOM-EA-CCSDT) calculations.\cite{Kucharski2001,
Bomble2004} Our EOM-EA-CCSDT calculations make use of the continuum orbital 
method:\cite{Stanton1999} A basis function with an exponent of 10$^{-10}$, 
which effectively does not interact with the rest of the basis set, is 
included in the calculation and two electrons are placed in the resulting 
orbital when solving the restricted HF (RHF) equations. In a subsequent 
EOM electron-excitation CCSDT (EOM-EE-CCSDT) calculation, one electron 
can then be excited from the continuum orbital into a target orbital 
allowing for the calculation of EOM-EA-CCSDT states using an EOM-EE-CCSDT 
implementation. 

Additionally, for HCl$^-$ and N$_2^-$, we compute the second moments of 
the electron density and the Dyson orbital of the anionic states at the 
EOM-EA-CCSD level. Dyson orbitals can be viewed as transition density 
matrices between neutral and electron-attached states and characterize 
electron attachment without invoking a mean-field approximation.\cite{
Oana2007,Jagau2016,Krylov2020} The frozen-core approximation is used in 
all CC calculations. For HCl, we also compute the energy of the anion 
at the unrestricted HF (UHF) level of theory and by applying Koopmans' 
theorem to the virtual orbitals of the neutral molecule.  
All CCSD, EOM-EA-CCSD, and UHF calculations were performed using the 
\textsc{Q-Chem} software, version 6.0.2.\cite{QChem5} All CCSDT and 
EOM-EA-CCSDT calculations were performed using the CFOUR software, 
version 2.1.\cite{cfour} 


\section{Results}
\subsection{Potential energy curve of HCl$^-$}

\begin{figure*}[hbt] \centering
\includegraphics[scale=0.97]{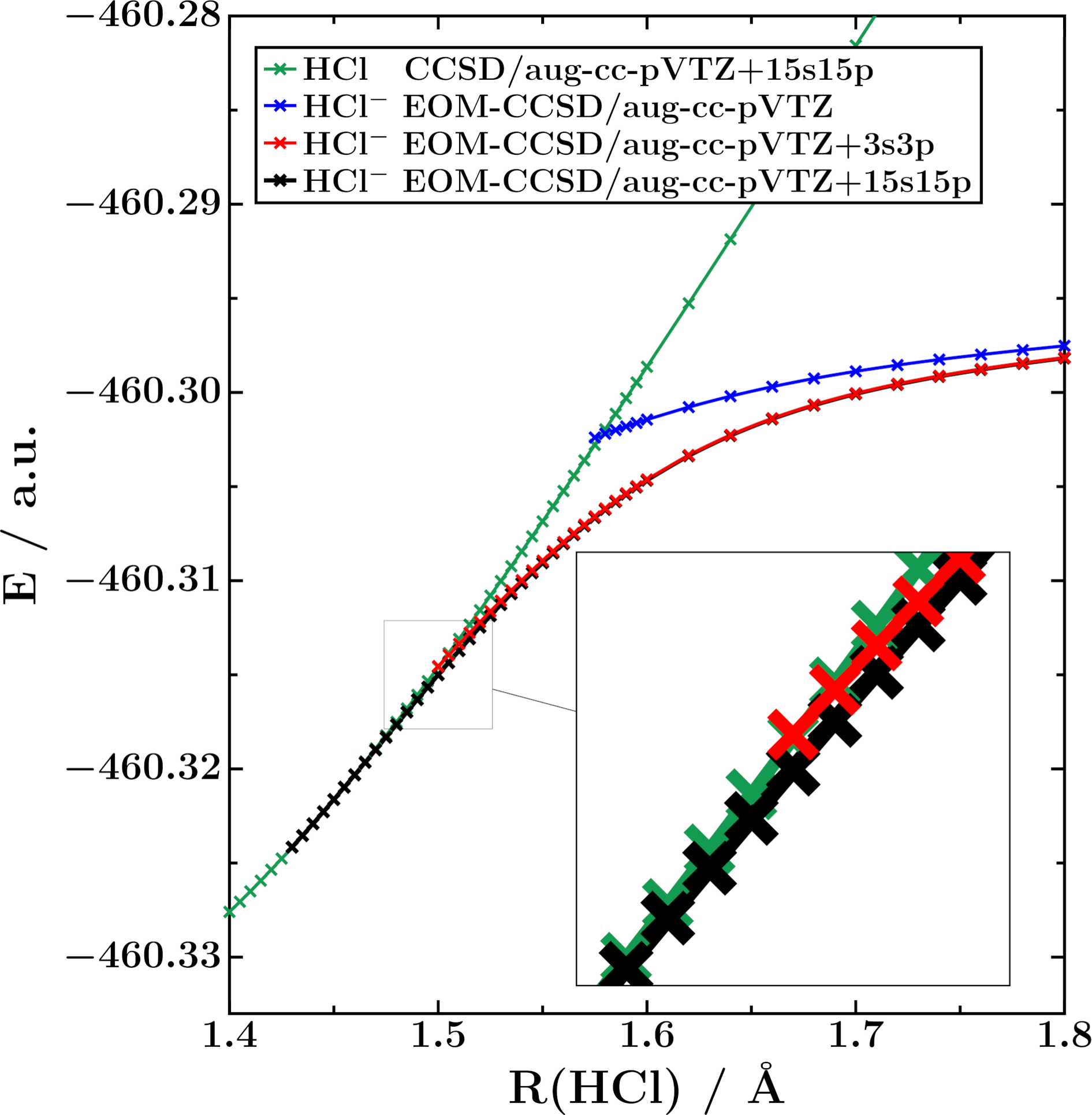}
\includegraphics[scale=1.0]{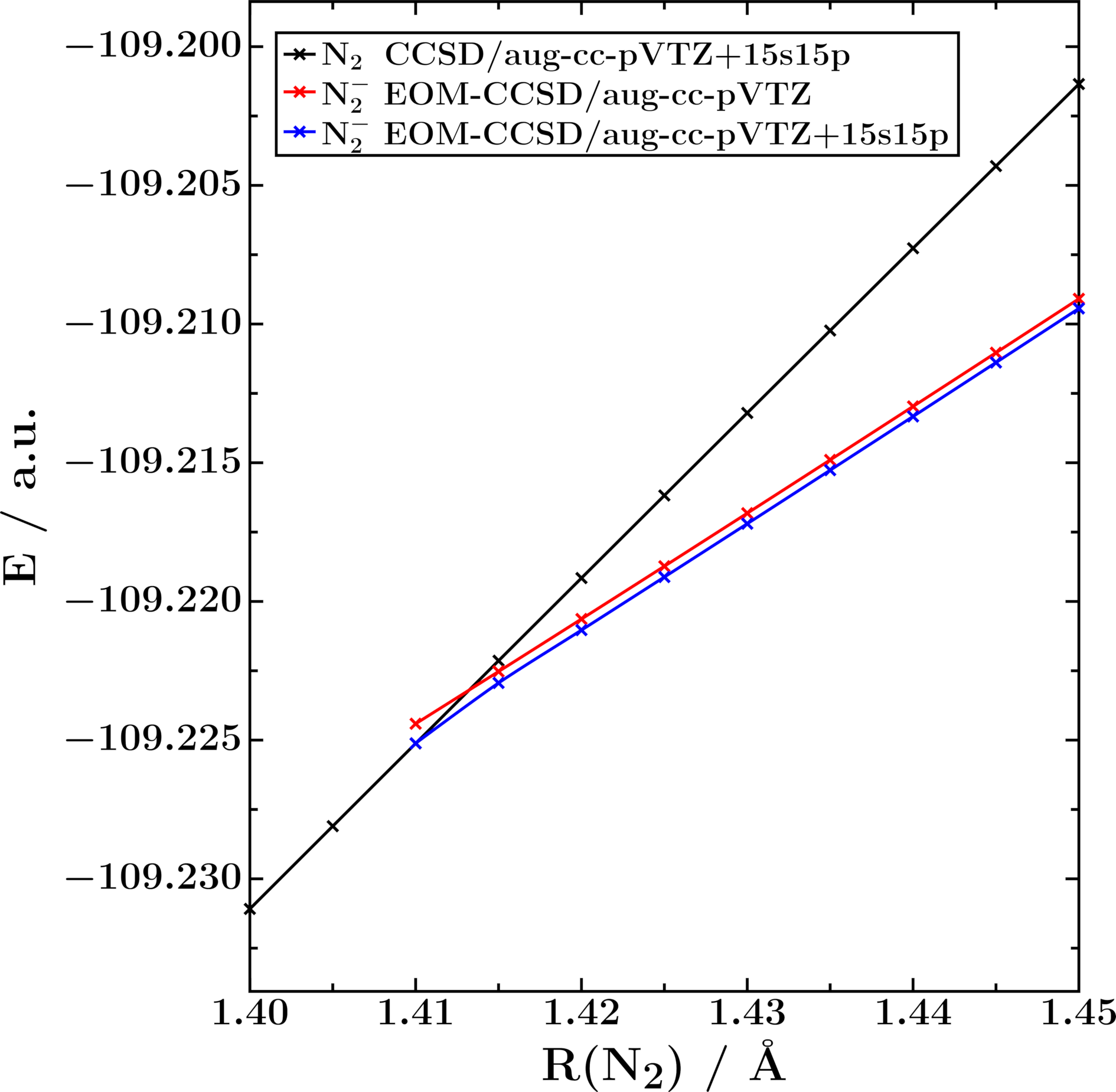}
\caption{Left: Potential energy curves of the $^1\Sigma^+$ ground state 
of HCl and the $^2\Sigma^+$ ground state of HCl$^-$ computed with CCSD 
and EOM-EA-CCSD, respectively, at $1.4 < $ R(HCl) $ < 1.8$ \AA. Right: Potential 
energy curves of the $^1\Sigma^+_\mathrm{g}$ ground state of N$_2$ and the $^2\Pi_\mathrm{g}$ 
ground state of N$_2^-$ computed with CCSD and EOM-EA-CCSD, respectively, 
at 1.4 $<$ R(NN) $ < 1.45$ \AA.}
\label{fig:hcl1}
\end{figure*}

In the left panel of Fig. \ref{fig:hcl1} we show potential energy curves 
of neutral HCl and of the $^2\Sigma^+$ ground state of HCl$^-$ computed 
using CCSD and EOM-EA-CCSD, respectively, and three different basis sets: 
aug-cc-pVTZ, aug-cc-pVTZ+3s3p and aug-cc-pVTZ+15s15p. To illustrate the 
special character of these potential energy curves, we show in the right 
panel of the same figure the corresponding curves for neutral N$_2$ and 
the $^2\Pi_\mathrm{g}$ ground state of N$_2^-$ computed with the same methods and 
basis sets. 

Fig. \ref{fig:hcl1} demonstrates that the inclusion of diffuse shells in 
the basis set has a much bigger effect on HCl$^-$ than on N$_2^-$. In the 
unmodified aug-cc-pVTZ basis, both anionic potential energy curves behave 
similarly, but upon the addition of diffuse shells the HCl$^-$ curve bends 
downwards and crosses the curve of the neutral molecule at a much shorter 
H-Cl distance. 

\begin{table} \centering
\caption{Crossing points (\AA) of the $^2\Sigma^+$ ground state of 
HCl$^-$ with the $^1\Sigma^+$ ground state of HCl computed with 
EOM-EA-CCSD, EOM-EA-CCSDT, $\Delta$HF, and Koopmans' theorem (KT) 
using basis sets with an increasing number of diffuse shells.}
\begin{tabular*}{0.7\textwidth}{@{\extracolsep{\fill}} l *4c}\hline
Basis set & KT & $\Delta$HF & CCSD & CCSDT \\ \hline
aug-cc-pVTZ        & 1.735 & 1.639 & 1.577 & 1.569 \\
aug-cc-pVTZ+3s3p   & 1.682 & 1.630 & 1.502 & 1.498 \\
aug-cc-pVTZ+5s5p   & 1.664 & 1.630 & 1.475 & 1.474 \\
aug-cc-pVTZ+10s10p & 1.642 & 1.630 & 1.445 & \\
aug-cc-pVTZ+15s15p & 1.631 & 1.630 & 1.434 & \\ \hline
\end{tabular*} \label{tab:hcl1}
\end{table}

The crossing points between the neutral and anionic HCl potential energy 
curves computed with different methods and basis sets can be found in 
Table \ref{tab:hcl1}. It is seen that this point moves by 0.143 \AA\ at 
the EOM-EA-CCSD level when going from the standard aug-cc-pVTZ basis to 
aug-cc-pVTZ+15s15p. In fact, we cannot assure that the value of 1.434 \AA\ 
computed in the largest basis is converged with respect to basis-set size 
because the inclusion of 15 diffuse s and p shells represents a technical 
limit of our implementation. In contrast, the crossing point between the 
N$_2$ and N$_2^-$ curves in the right panel of Fig. \ref{fig:hcl1} is 
nearly invariant and only moves from 1.413 \AA\ to 1.412 \AA\ upon the 
inclusion of 15 diffuse s and p shells. We also note that the basis-set 
dependence of the HCl$^-$ curve is only significant near the crossing 
point. At H-Cl bond lengths of 1.8 \AA, the differences between the basis 
sets are negligible.  

To confirm the validity of the EOM-EA-CCSD approximation for the HCl$^-$ 
anion, we conducted EOM-EA-CCSDT calculations, which are documented in 
Table \ref{tab:hcl1} as well. This illustrates that the crossing point 
moves to slightly shorter bond distances as compared to EOM-EA-CCSD but 
the bending effect remains similar in magnitude. However, due to technical 
limitations, we could not run EOM-EA-CCSDT calculations with more than 
5 diffuse s and p shells. For N$_2^-$, differences between EOM-EA-CCSDT 
and EOM-EA-CCSD are very similar to those observed for HCl$^-$, which 
corroborates that the bending feature observed for HCl$^-$ is well 
captured by EOM-EA-CCSD and not related to an insufficient treatment 
of electron correlation.

\begin{figure}[hbt] \centering
\includegraphics{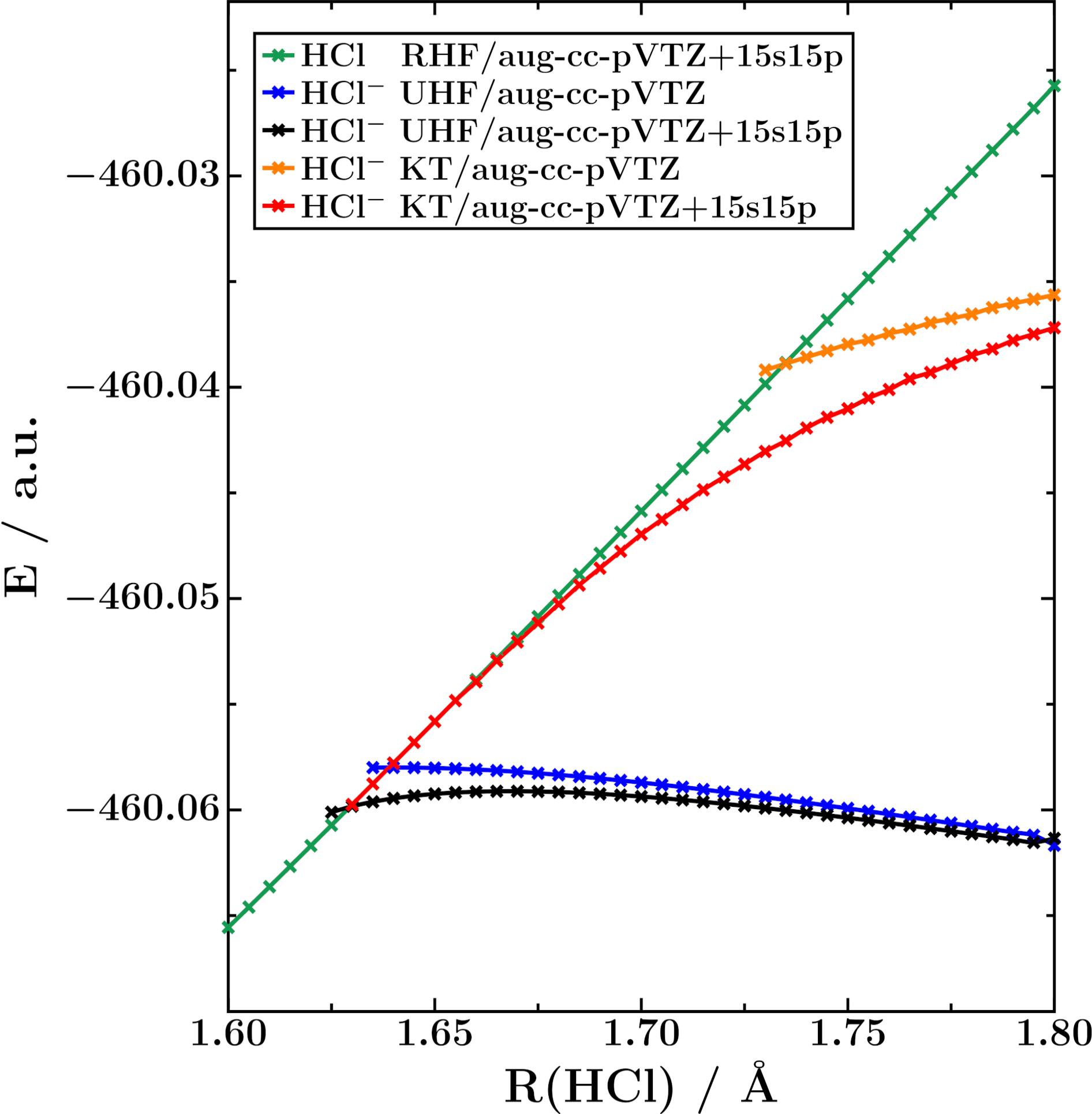}
\caption{Potential energy curves of the $^1\Sigma^+$ ground state 
of HCl and the $^2\Sigma^+$ ground state of HCl$^-$ computed with HF 
theory and Koopmans' theorem at 1.6 $ < $ R(HCl) $ < 1.8 $ \AA.}
\label{fig:hcl2}
\end{figure}

To further investigate the nature of HCl$^-$, we also conducted UHF 
calculations on this anion and we also applied Koopmans' theorem to 
the lowest-lying virtual orbital of neutral HCl. The resulting potential 
energy curves are shown in Fig. \ref{fig:hcl2}, the corresponding 
crossing points are available in Table \ref{tab:hcl2}. It is seen 
that UHF does not recover the bending effect and basis-set dependence 
visible in the EOM-CC potential energy curves from Fig. \ref{fig:hcl1}, 
while Koopmans' theorem does. At HF level of theory, the inclusion of 
15 diffuse s and p shells in the basis set moves the crossing point 
by only 0.009 \AA, whereas that value amounts to 0.104 \AA\ with 
Koopmans' theorem. 

We add that there is a region of ca. 0.2 \AA, from 1.63 {\AA} to 1.434 \AA, 
where HCl$^-$ is only bound at correlated levels of theory. At these H-Cl 
distances, the energy of the anionic UHF potential curve lies above the 
RHF neutral potential curve, while the EOM-EA-CCSD anionic potential 
curve lies below the CCSD neutral potential curve. Also this behaviour 
is qualitatively different from N$_2^-$, which is bound already at shorter 
distances at the HF level of theory (R(NN) $\approx$ 1.33 \AA) than at the 
EOM-EA-CCSD level of theory (R(NN) $\approx$ 1.41 \AA).\cite{Jagau2017}

We note that the dipole moment of HCl, computed with CCSD/aug-cc-pVTZ, 
amounts to 1.48 D and 1.24 D at R(H-Cl)=1.8 {\AA} and 1.434 \AA, 
respectively. This is substantially below the critical value needed 
to form a dipole-bound anion,\cite{Simons2008} meaning that the 
polarity of HCl is not the critical difference to N$_2$. However, 
the same bending effect shown for HCl$^-$ in Figs. \ref{fig:hcl1} 
and \ref{fig:hcl2} has also been observed in EOM-EA-CCSD calculations 
along the bending coordinate of CO$_2^-$ and connected to s-wave 
scattering.\cite{Sommerfeld2003} A further interesting parallel can 
be drawn to correlation-bound anions such as C$_6$F$_6^-$\cite{Voora2014} 
and C$_{60}^-$:\cite{Voora2013} Also for these electronic states, a 
description based on a UHF wave function of the anion yields poor 
results, while EOM-EA-CC calculations based on an orbital manifold 
optimized for the neutral molecule perform much better.


\subsection{Second moment of the electron density of HCl$^-$}

\begin{table*} \centering
\caption{Second moment of the electron density ({\AA}$^2$) of the 
$^2\Sigma^+$ ground state of HCl$^-$ computed at various internuclear 
distances with EOM-EA-CCSD using basis sets with an increasing number 
of diffuse shells}
\begin{tabular*}{\textwidth}{@{\extracolsep{\fill}} l *6r} \hline
Basis set & Crossing point & 1.45 \AA & 1.50 \AA & 1.55 \AA & 1.60 \AA & 1.80 \AA \\ \hline
aug-cc-pVTZ        & 20.3 \; (1.577 \AA) &  &  &  & 19.8 & 17.5 \\
aug-cc-pVTZ+3s3p   & 71.4 \; (1.502 \AA) &  &  & 48.6 & 31.4 & 18.0 \\
aug-cc-pVTZ+5s5p   & 250.1 \; (1.475 \AA) &  & 170.9 & 61.8 & 32.3 & 18.0 \\
aug-cc-pVTZ+10s10p & 6897.8 \; (1.445 \AA) & 5507.7 & 255.2 & 62.5 & 32.4 & 18.6 \\
aug-cc-pVTZ+15s15p & 647195.1 \; (1.434 \AA) & 449047.1 & 255.7 & 62.5 & 32.4 & 18.4 \\ \hline
\end{tabular*}
\label{tab:hcl2}
\end{table*}

To further characterise the HCl$^-$ anion, we computed second moments of 
the electron density ($\langle r^2 \rangle$) with EOM-EA-CCSD and different 
basis sets. The results are reported in Table \ref{tab:hcl2}. At 1.8 \AA\ 
where HCl$^-$ is in all basis sets bound in terms of Koopmans' theorem, 
$\langle r^2 \rangle$ is around 18 \AA$^2$. Increasing the number of 
diffuse shells has only a small impact on $\langle r^2 \rangle$ at this 
bond length. At 1.60 \AA, where HCl$^-$ is bound at the EOM-EA-CCSD level 
but already unbound at the HF level, the basis-set dependence of $\langle 
r^2 \rangle$ is already more pronounced; 3 additional diffuse s and 
p shells on top of aug-cc-pVTZ lead to an increase by 50\%. However, 
even more diffuse shells only become relevant at shorter distances: At 
1.50 \AA, the value of $\langle r^2 \rangle$ has increased to 255 \AA$^2$ 
and 10 diffuse s and p shells are need to capture it. At 1.434 \AA\, i.e., 
the crossing point in our largest basis set (aug-cc-pVTZ+15s15p), we obtain 
a value of 647,000 \AA$^2$ for $\langle r^2 \rangle$, which corresponds to 
an average distance of ca. 800 \AA. Moreover, we cannot say whether this 
value is actually converged with respect to basis-set size.  

\begin{table} \centering
\caption{Second moment of the electron density ({\AA}$^2$) of the 
$^2\Pi_\mathrm{g}$ ground state of N$_2^-$ computed at various internuclear 
distances with EOM-EA-CCSD using basis sets with an increasing number 
of diffuse shells}
\begin{tabular*}{0.7\textwidth}{@{\extracolsep{\fill}} l *3r} \hline
Basis set & Crossing point & 1.6 \AA & 1.8 \AA \\ \hline
aug-cc-pVTZ        & 17.3 & 19.5 & 22.1 \\
aug-cc-pVTZ+3s3p   & 17.7 & 19.5 & 22.1 \\
aug-cc-pVTZ+5s5p   & 17.8 & 19.5 & 22.1 \\
aug-cc-pVTZ+10s10p & 17.9 & 19.5 & 22.1 \\
aug-cc-pVTZ+15s15p & 17.7 & 19.5 & 22.1 \\ \hline
\end{tabular*}
\label{tab:n2}
\end{table}

The explosion of $\langle r^2 \rangle$ by a factor of more than $2 \cdot 
10^5$ over a range of less than 0.2 \AA\ explains the pronounced basis-set 
dependence of the HCl$^-$ potential energy curve in Fig. \ref{fig:hcl1}, 
but it is by no means typical of molecular anions. As a counterexample, 
we present in Table \ref{tab:n2} $\langle r^2 \rangle$ values for N$_2^-$. 
For this anion, the second moment is nearly invariant with respect to 
basis-set changes and, in fact, decreases somewhat when approaching the 
crossing point. 

\subsection{Dyson orbital of HCl$^-$} 

\begin{figure*}[bht] \centering
\includegraphics[scale=0.75]{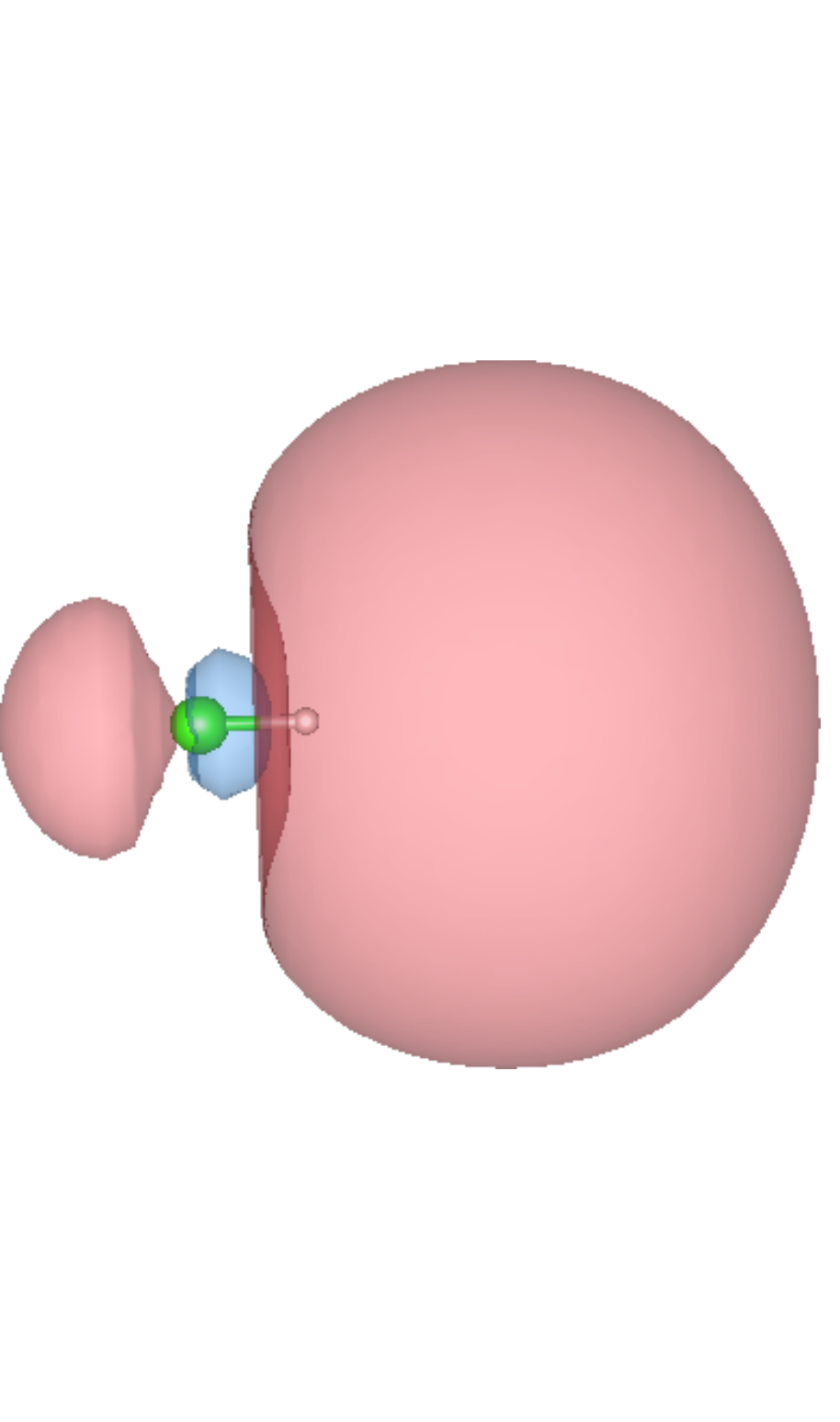}\hspace{0.5cm}
\includegraphics[scale=0.75]{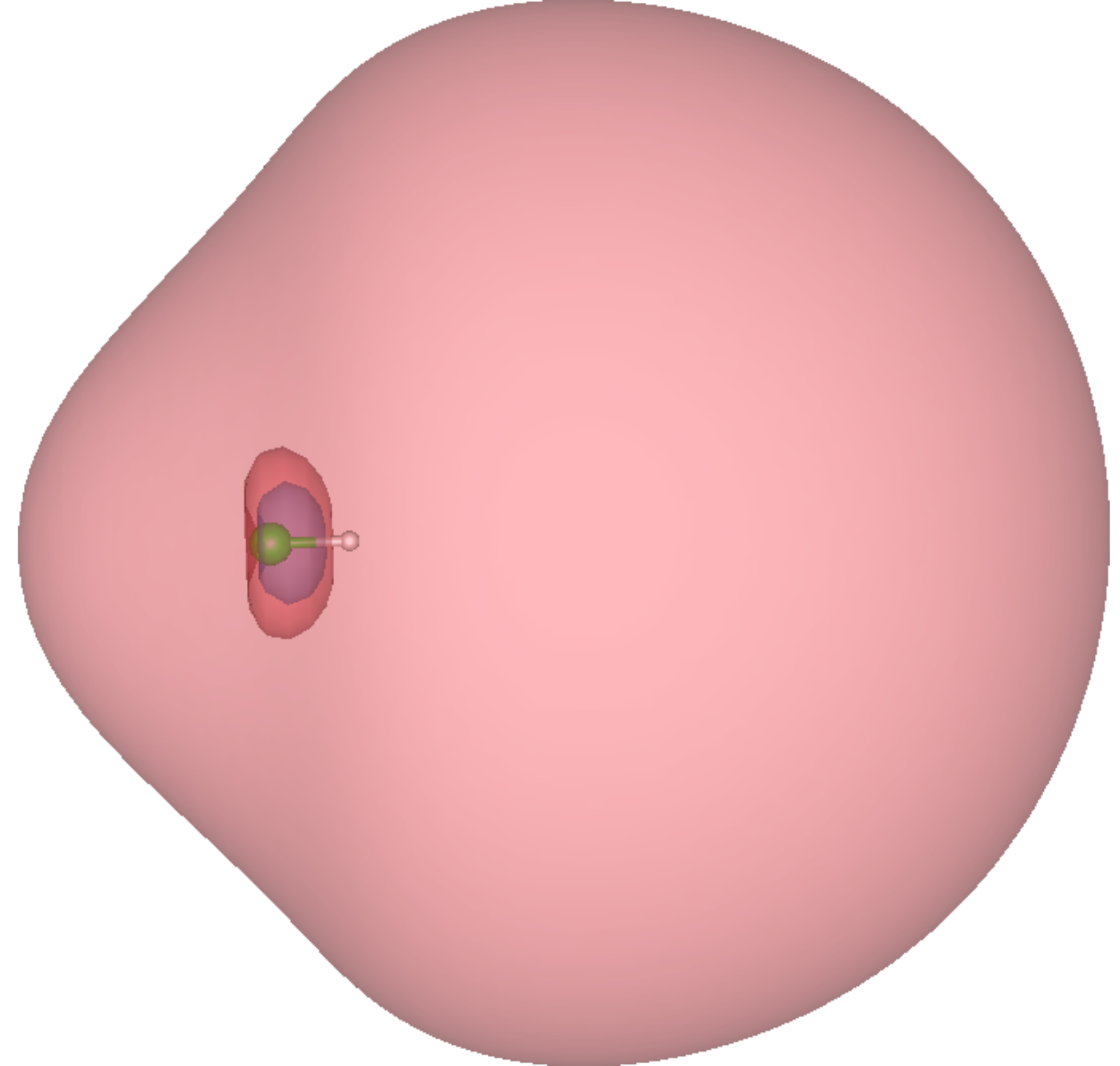} \hspace{3cm}
\includegraphics[scale=0.75]{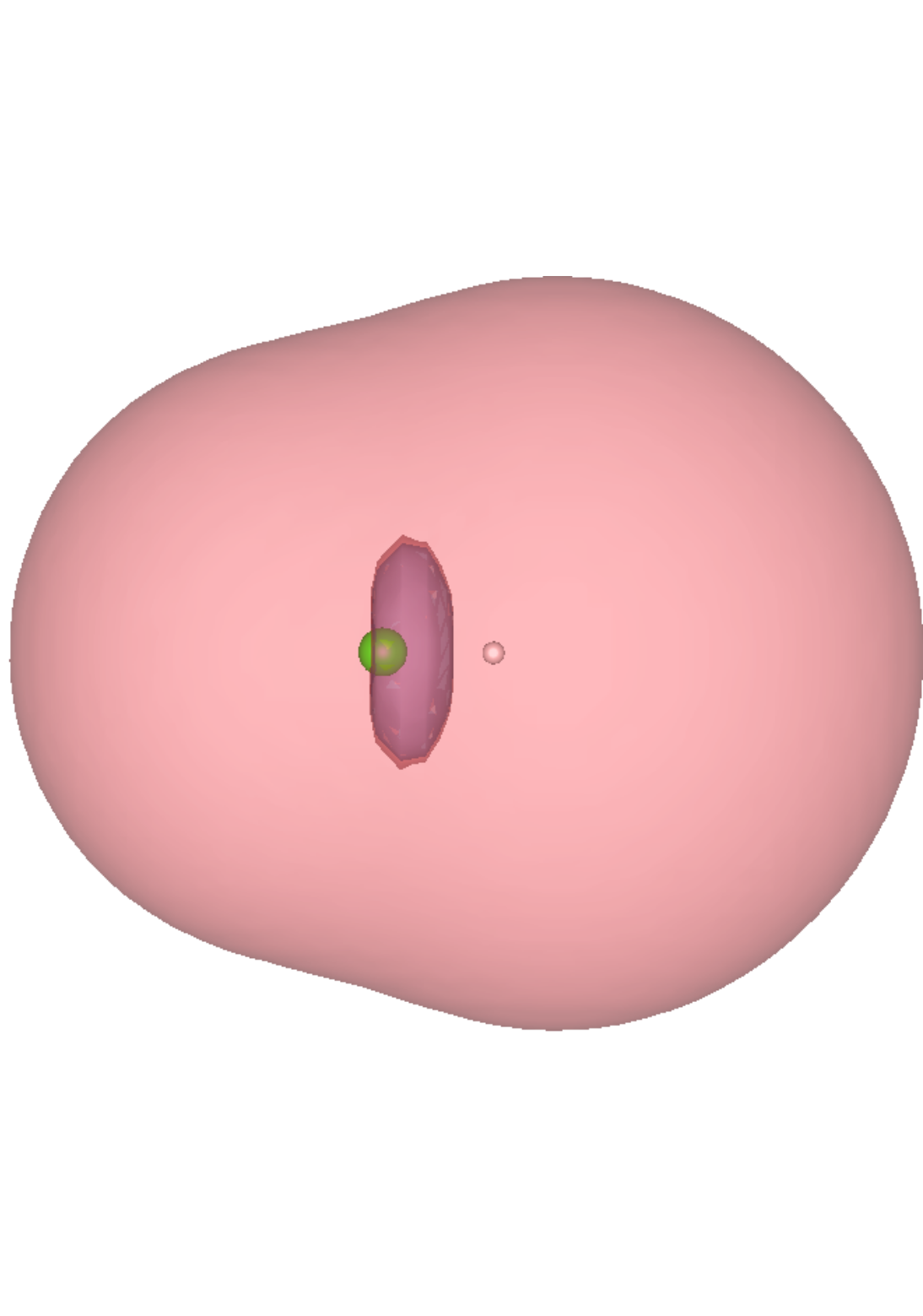}
\caption{Dyson orbital of the $^2\Sigma^+$ ground state of HCl$^-$ 
computed with EOM-EA-CCSD/aug-cc-pVTZ+15s15p. Left: R(HCl) = 1.434 \AA, 
isovalue = $5 \cdot 10^{-5}$; middle: R(HCl) = 1.434 \AA, isovalue = 
$3 \cdot 10^{-5}$; right: R(HCl) = 1.8 \AA, isovalue =  $1 \cdot 10^{-3}$}
\label{fig:hcldy}
\end{figure*}

\begin{figure} \centering
\includegraphics{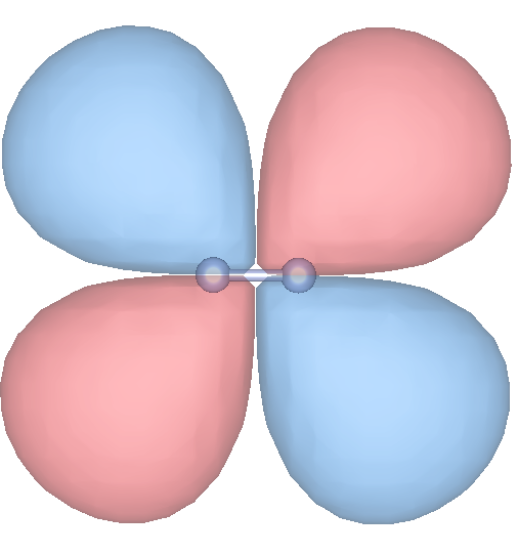} \hspace{2cm}
\includegraphics{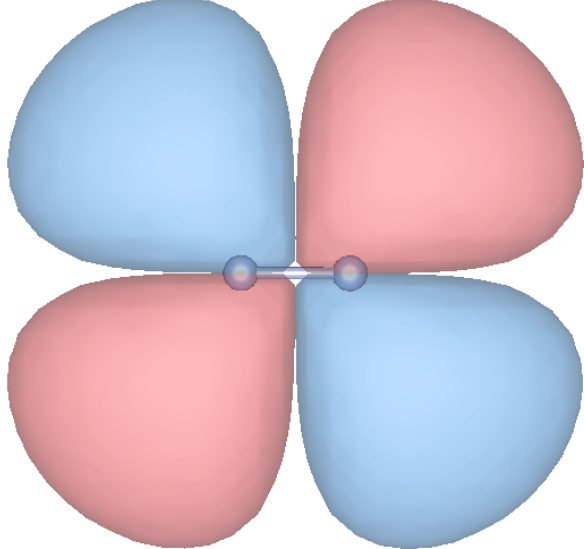}
\caption{Dyson orbital of the $^2\Pi_\mathrm{g}$ ground state of N$_2^-$ 
computed with EOM-EA-CCSD/aug-cc-pVTZ+15s15p and plotted at an isovalue 
of 0.001 at internuclear distances of 1.413 {\AA} (left) and 1.800 {\AA} 
(right)}
\label{fig:n2dy}
\end{figure}

The extremely diffuse nature of the EOM-EA-CCSD wave function of HCl$^-$ 
near the crossing point with the neutral potential curve is also captured 
by the corresponding Dyson orbital, which is displayed in Fig. 
\ref{fig:hcldy}. Plotted at R(H-Cl) = 1.434 \AA\ and an isovalue of $5 
\times 10^{-5}$ (left panel of Fig.\ref{fig:hcldy}), the orbital resembles 
a $\upsigma^*$ orbital between hydrogen s and chlorine p atomic orbitals 
with a node in the centre of the bond. On reducing the isovalue to $3 
\times 10^{-5}$ (middle panel of Fig.\ref{fig:hcldy}), the character of 
the orbital changes and it becomes more akin to a $\upsigma$ orbital with 
a larger amplitude on the side of the hydrogen atom. At R(H-Cl) = 1.8 \AA\ 
(right panel of Fig. \ref{fig:hcldy}), the $\upsigma$-character of the Dyson 
orbital is preserved but its spatial extent is considerably smaller consistent 
with the trend seen in $\langle r^2 \rangle$ in Table \ref{tab:hcl2}. 

Similar to the second moment, the behaviour of the Dyson orbital of 
HCl$^-$ is by no means typical of molecular anions as illustrated by 
Fig. \ref{fig:n2dy}, which shows the Dyson orbital of N$_2^-$. In contrast 
to HCl$^-$, the Dyson orbital of N$_2^-$ remains compact and of $\uppi^*$ 
character independent of the isovalue and also when the N-N bond length 
is decreased from 1.8 \AA\ to 1.413 \AA, where the potential curves of 
N$_2^-$ and N$_2$ cross. In fact, this orbital becomes slightly less 
diffuse when the bond becomes shorter. With non-Hermitian techniques, it 
is possible to evaluate a complex-valued Dyson orbital in the unbound 
region at R(NN) $ < 1.413$ \AA\ as well;\cite{Jagau2016} the real part of 
such an orbital looks very similar to those displayed in Fig. \ref{fig:n2dy}.

The Dyson orbital of HCl$^-$ shown in Fig. \ref{fig:hcldy} is, however, 
similar to what one observes for correlation-bound anions.\cite{Voora2013,
Voora2014,Jordan2017,Paran2024} In C$_6$F$_6^-$, for example, the EOM-EA-CC 
natural orbital hosting the excess electron has a spatial extent much 
larger than the nuclear framework.\cite{Voora2014} 

\subsection{Anions of H$_2$} 

\begin{figure*} \centering
\includegraphics{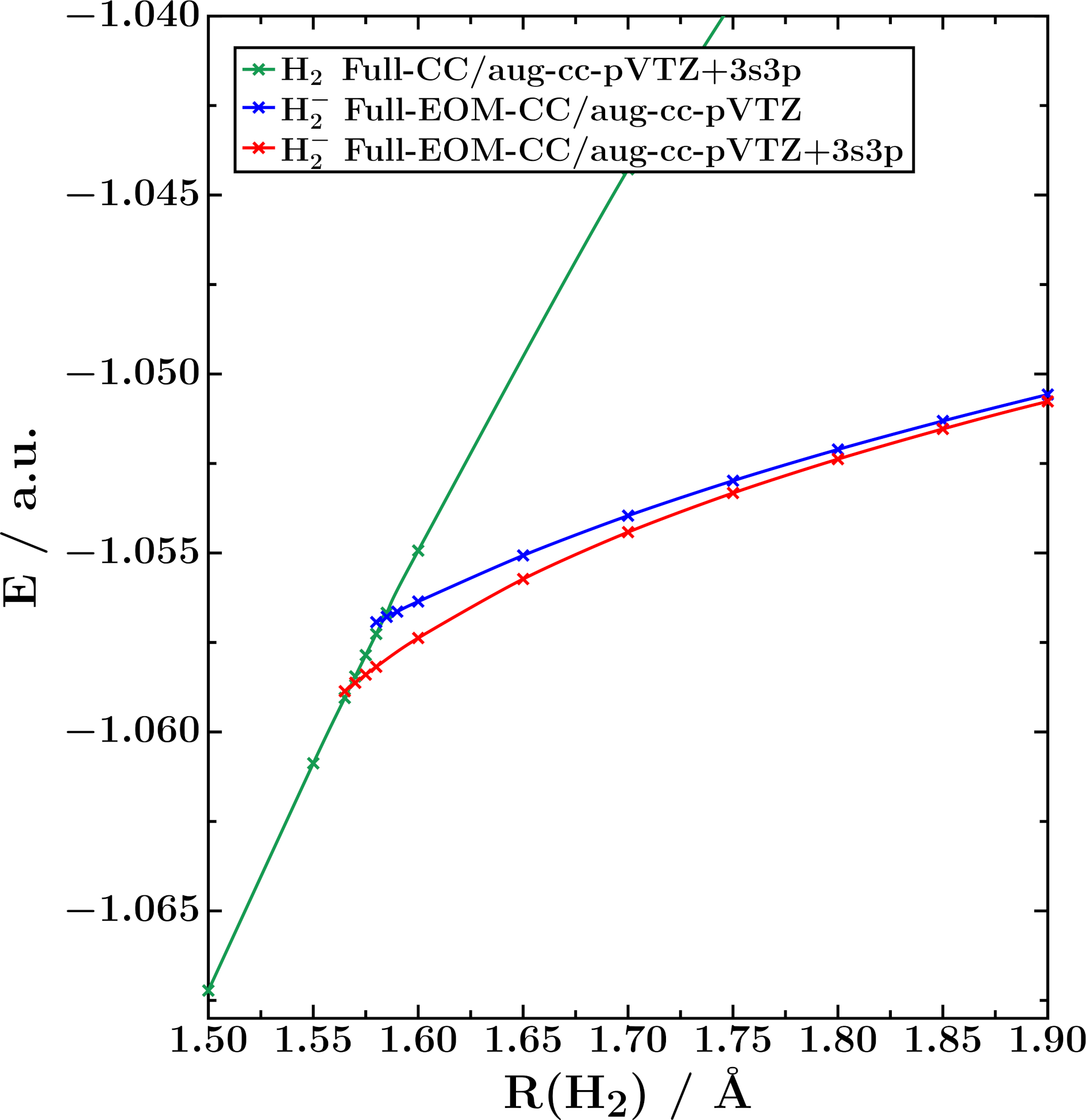}
\includegraphics{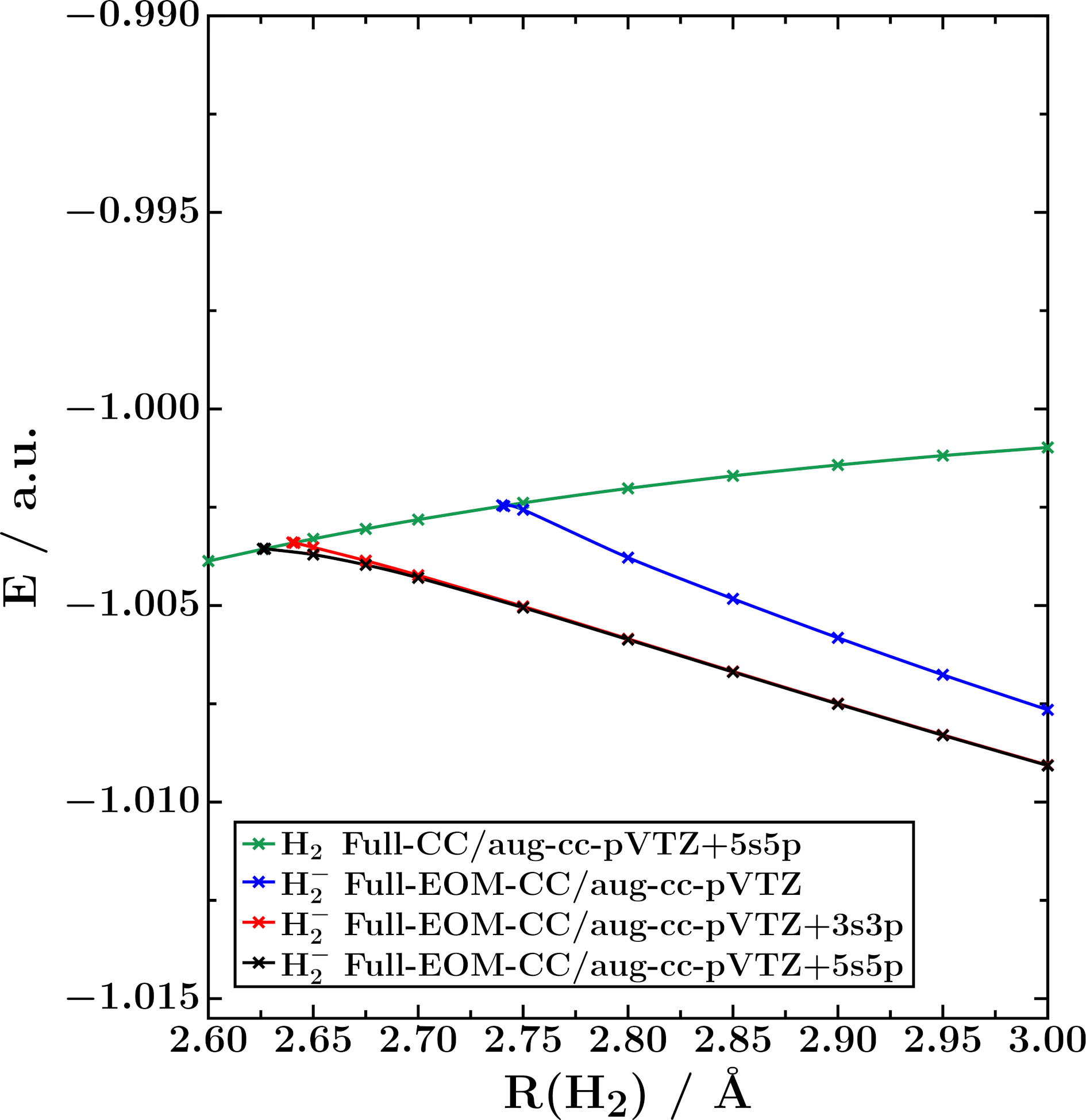}
\caption{Potential energy curves of the $^2\Sigma^+_u$ (left) and 
$^2\Sigma^+_g$ (right) states of H$_2^-$ computed with full EOM-EA-CC.
The $^1\Sigma^+_g$ ground state of H$_2$ is also shown.}
\label{fig:h2}
\end{figure*}

There are two bound doublet states of H$_2^-$: a $^2\Sigma^+_\mathrm{u}$ state, 
which is bound at bond lengths above ca. 1.6 \AA, and a $^2\Sigma^+_\mathrm{g}$ 
state, which is bound at bond lengths above ca. 2.7 \AA. The configuration 
of the first state is $(\upsigma_\mathrm{g})^2(\upsigma_\mathrm{u})^1$, while that of the second 
state is $(\upsigma_\mathrm{g})^1(\upsigma_\mathrm{u})^2$. The first state turns into a shape 
resonance at shorter bond lengths, whereas the second state is first a 
Feshbach resonance and then becomes a shape resonance as soon as the 
$^3\Sigma_\mathrm{u}^+$ state of H$_2$ is higher in energy.\cite{White2017} 
As H$_2^-$ has only three electrons, both states can be treated 
exactly within a given basis set by means of EOM-EA-CCSDT calculations. 
Particularly relevant in the context of the present work is that the 
electron emitted in the decay of the $^2\Sigma^+_\mathrm{g}$ state of H$_2^-$ 
has to have zero angular momentum, meaning it represents an s-wave. 

The potential curves of both anionic states computed with full EOM-EA-CC 
in different basis sets are shown in Fig. \ref{fig:h2}; the corresponding 
crossing points with the potential energy curve of neutral H$_2$ are 
summarised in Table \ref{tab:h2}. Fig. \ref{fig:h2} shows that none of 
the two H$_2^-$ states shows the bending effect observed in HCl$^-$. 
Also, the position of the crossing point between the neutral and anionic 
potential curves depends much less on the basis set than in the case 
of HCl$^-$. 

For the $^2\Sigma^+_\mathrm{u}$ state, the position of the crossing point is 
converged in the aug-cc-pVTZ+3s3p basis, where it differs by only 
0.02 \AA\ from the position in the aug-cc-pVTZ basis. The crossing 
point of the $^2\Sigma^+_\mathrm{g}$ state does vary more than that of the 
ungerade state when diffuse shells are added to the basis, but the 
position is converged within the aug-cc-pVTZ+5s5p basis and additional 
diffuse shells make no impact. We note that at much longer bond 
distances of 4.0 and 5.0 \AA\ the potential curves obtained in the 
aug-cc-pVTZ and aug-cc-pVTZ+5s5p basis sets continue to run parallel. 

This indicates that the aug-cc-pVTZ basis is insufficient to describe 
the $^2\Sigma^+_\mathrm{g}$ state at any bond length and the discrepancy in the 
position of the crossing points is not reflective of the bending effect 
observed in HCl$^-$, for which diffuse shells make very little impact 
at long bond distances. We finally note that the differences between 
EOM-EA-CCSD and EOM-EA-CCSDT are for both states of H$_2^-$ similar 
to those observed for HCl$^-$ and N$_2^-$. 

\begin{table} \centering
\caption{Crossing points ({\AA}) of the $^2\Sigma^+_u$ and $^2\Sigma^+_g$ 
states of H$_2^-$ with the $^1\Sigma^+_g$ ground state of H$_2$ computed 
with EOM-EA-CCSD and EOM-EA-CCSDT using basis sets with an increasing 
number of diffuse shells. For EOM-EA-CCSD calculations on the ungerade 
state, we use the $^3\Sigma^+_u$ state of HCl as reference to enable a 
description as singly-excited state.}
\begin{tabular*}{0.7\textwidth}{@{\extracolsep{\fill}} l *4r} \hline
 & \multicolumn{2}{c}{$^2\Sigma^+_u$} & \multicolumn{2}{c}{$^2\Sigma^+_g$} \\ 
\cline{2-3} \cline{4-5}
Basis set & CCSD & CCSDT & CCSD & CCSDT \\ \hline
aug-cc-pVTZ        & 1.613 & 1.585 & 2.760 & 2.741 \\
aug-cc-pVTZ+3s3p   & 1.594 & 1.570 & 2.672 & 2.641 \\
aug-cc-pVTZ+5s5p   & 1.593 &  & 2.668 & 2.627 \\
aug-cc-pVTZ+10s10p & 1.593 &  & 2.668 & \\
aug-cc-pVTZ+15s15p & 1.593 &  & 2.668 & \\ \hline
\end{tabular*}
\label{tab:h2}
\end{table}

\subsection{Pyrrole anion}

\begin{figure}
\centering
\includegraphics{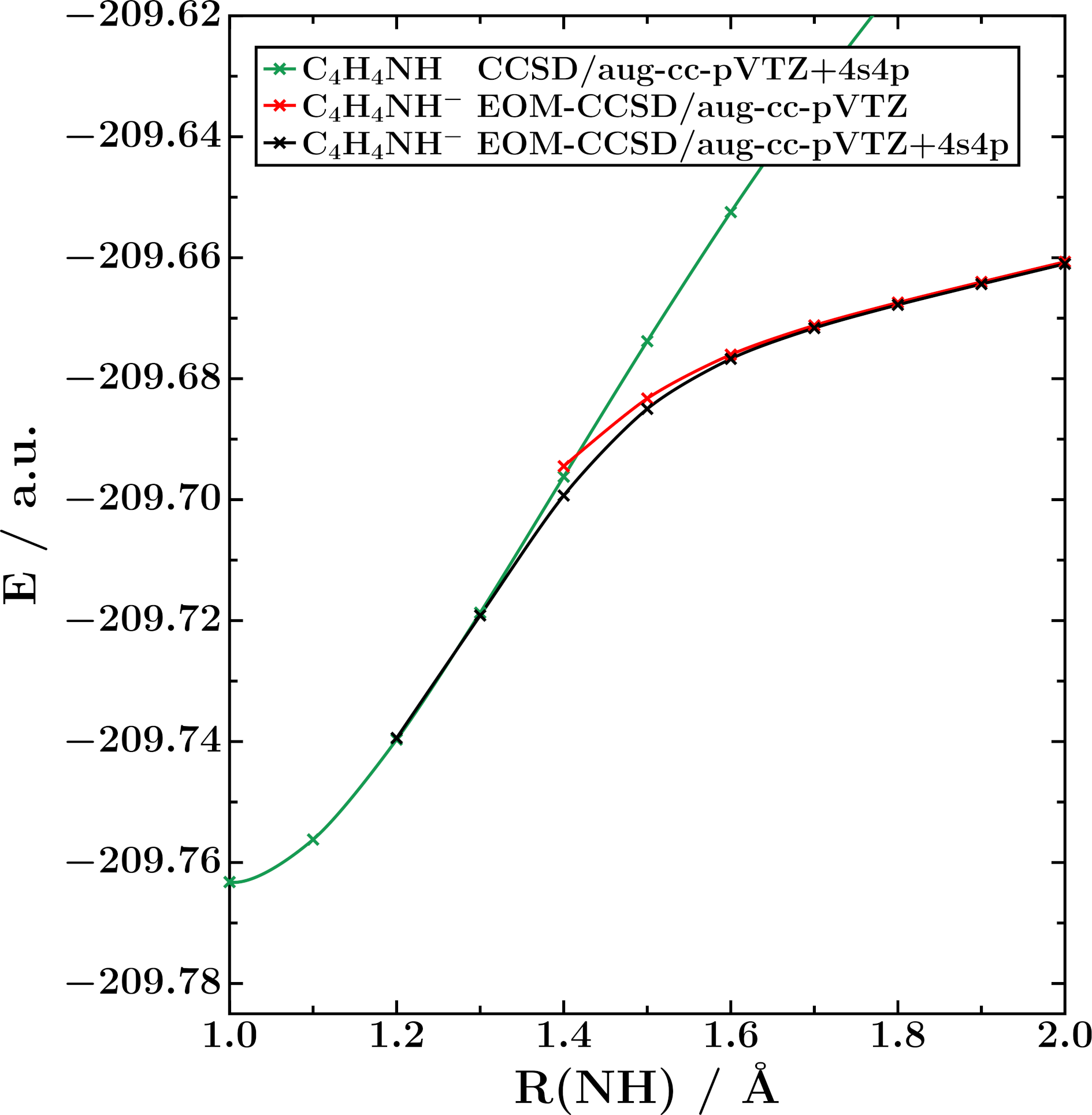}
\caption{Potential energy curves of the $^1$A$_1$ ground state of 
pyrrole and the $^2$A$_1$ state of the pyrrole anion computed with 
CCSD and EOM-EA-CCSD, respectively, as a function of N-H bond length 
at 1.0 $ < $ R(NH) $ < $ 2.0 \AA.}
\label{fig:pyrrole}
\end{figure}

To demonstrate the relevance of our findings beyond diatomic molecules, 
we investigated the $^2$A$_1$ state of the pyrrole anion. Pyrrole does 
not support a bound anion at its equilibrium structure but has a rich 
electron-induced chemistry.\cite{Ragesh2022,Mukherjee2022,Modelli2004,
Oliveira2010,Pshenichnyuk2019} While the nature of two $\uppi^*$ resonances 
is not debated, it is less clear if the low-lying totally symmetric anionic 
state, which is also present in pyrrole, should be interpreted as virtual 
state or rather as $\upsigma^*$ resonance.\cite{Pshenichnyuk2019,Ragesh2022,
Mukherjee2022} Independent of the nature of the latter state, there are 
indications that non-adiabatic transitions\cite{Chatterjee2023} between 
the $\uppi^*$ resonances and the totally symmetric state mediate dissociative 
electron attachment, which has implications for similar processes in more 
complex molecules.\cite{Hotop2003,Simons2008} 

For a comprehensive investigation of the pyrrole anion, the full-dimensional 
potential energy surface would need to be studied. By means of analytic 
gradient techniques,\cite{Benda2017,Benda2018a,Benda2018b} this is, in 
principle, possible also in the unbound regions as recent applications 
illustrated.\cite{Mukherjee2022} However, here we concentrate on the 
bound regions of the potential energy surface and furthermore limit 
ourselves to the N-H stretching coordinate. 

As Figure \ref{fig:pyrrole} illustrates, the bound part of the potential 
energy curve of the $^2$A$_1$ state of the pyrrole anion shows the same 
bending effect and basis-set dependence that we documented for HCl$^-$ 
in Fig. \ref{fig:hcl1} and that was previously observed in CO$_2^-$.\cite{
Sommerfeld2003} At the EOM-EA-CCSD level, the $^2$A$_1$ state is bound 
at R(NH) $ >  $ 1.2 \AA\ in the aug-cc-pVTZ+4s4p basis, but only at R(NH) 
$ > $ 1.4 \AA\ in the unmodified aug-cc-pVTZ basis. Similar to HCl$^-$, the 
diffuse functions make no substantial impact at larger N-H distances of 
around 2 \AA. This strongly suggests that the $^2$A$_1$ state turns into 
a virtual state in the unbound region and not a $\upsigma^*$ resonance. 

\section{Conclusions}

We have investigated the bound anions of HCl, pyrrole, N$_2$, and H$_2$ 
using the EOM-EA-CCSD and EOM-EA-CCSDT methods at molecular structures 
close to where they become unbound. The $^2\Sigma^+$ state of HCl$^-$ 
and the $^2$A$_1$ state of pyrrole anion, both of which are connected 
to s-wave scattering, show a number of features that are not present 
in N$_2^-$ and H$_2^-$: Close to the crossing point with the parent 
neutral state, the spatial extent of the anionic wave function increases 
exponentially, which renders the potential energy curve extremely sensitive 
to diffuse functions in the basis set. For HCl$^-$, the inclusion of 15 
diffuse s and p shells with exponents down to $10^{-7}$ on top of the 
aug-cc-pVTZ basis set yields a value of 647,000 \AA$^2$ for the second 
moment of the electron density and moves the crossing point with the 
potential energy curve of neutral HCl by 0.143 \AA\ compared to aug-cc-pVTZ. 
Likely, these values are not yet converged with respect to the size of 
the basis set. In contrast, the $^2\Pi_\mathrm{g}$ state of N$_2^-$ and the 
$^2\Sigma_\mathrm{u}^+$ state of H$_2^-$, which are associated with d-wave 
scattering and p-wave scattering, respectively, are not sensitive 
to diffuse basis functions as their spatial extent does not change 
by a lot along the potential energy curve. Interestingly, the 
totally-symmetric $^2\Sigma_\mathrm{g}^+$ state of H$_2^-$, which is subject 
to a two-electron decay process past the crossing point, also does 
not show the features observed in HCl$^-$ and the anion of pyrrole. 

We note that the sensitivity of the potential energy curve towards 
diffuse basis functions was also observed for CO$_2^-$,\cite{Sommerfeld2003} 
another state associated with s-wave scattering. The unusual character 
of the HCl$^-$ anion is also visible in the Dyson orbital, which, close 
to the crossing point, acquires an extremely diffuse character that is 
akin to correlation-bound anions of, for example, C$_{60}$ or C$_6$F$_6$, 
which have been related to s-wave scattering as well.\cite{Voora2013,
Voora2014,Jordan2017}

Importantly, the non-totally symmetric states of H$_2^-$ and N$_2^-$, 
which turn into electronic shape resonances at shorter bond lengths, 
can be treated by means of non-Hermitian techniques such as complex 
basis functions or complex absorbing potentials in the unbound 
region.\cite{Jagau2014} The same applies to the totally-symmetric 
$^2\Sigma^+_\mathrm{g}$ state of H$_2^-$, which turns into a Feshbach 
resonance.\cite{White2017} 

In contrast, the totally-symmetric anions of HCl, CO$_2$, pyrrole, 
and similar molecules likely should not be considered as electronic 
resonances in the unbound region. Instead, the term ``virtual state'' 
has been coined for them. Whether these virtual states are amenable 
to a treatment in terms of complex absorbing potentials or complex 
basis functions, or how to integrate them by other techniques into 
bound-state electronic-structure theory, is not clear at present 
and we hope that our work stimulates further research in this direction. 


\section*{Conflicts of interest}
There are no conflicts to declare.

\section*{Acknowledgements}
R.E.M. acknowledges the European Union for mediating a bilateral exchange 
from the University of Durham to KU Leuven as a part of the Erasmus+ program.
This work has been supported by a Marie Sk\l{}odowska-Curie Actions fellowship 
to M.A. (Grant Agreement No. 101062717). T.-C.J. gratefully acknowledges funding 
from the European Research Council (ERC) under the European Union’s Horizon 2020 
research and innovation program (Grant Agreement No. 851766) and the KU Leuven 
internal funds (Grant No. C14/22/083).

\bibliography{s-waves} 
\bibliographystyle{rsc} 

\end{document}